\documentclass[twocolumn,twocolappendix]{aastex63}
\graphicspath{{./Figs/}{./png/}}
\usepackage{amsmath}
\let\tablenum\relax
\usepackage{siunitx}
\usepackage{multirow}

\newcommand{\K}{\,{\rm K}}
\newcommand{\G}{\,{\rm G}}

\newcommand{\nG}{\,{\rm nG}}
\newcommand{\g}{\,{\rm g}}
\newcommand{\cm}{\,{\rm cm}}
\newcommand{\kpc}{\,{\rm kpc}}
\newcommand{\Mpc}{\,{\rm Mpc}}
\newcommand{\cMpch}{\,h^{-1}{\rm cMpc}}

\newcommand{\Mpch}{\,h^{-1}{\rm Mpc}}
\newcommand{\s}{\,{\rm s}}
\newcommand{\cMpc}{\,{\rm cMpc}}

\newcommand{\ckpch}{\,h^{-1}{\rm ckpc}}

\newcommand{\yr}{\,{\rm yr}}

\shorttitle{Primordial Magnetic fields in the intergalactic medium}
\shortauthors{Mtchedlidze et al.}
\graphicspath{{./}{figures/}}

\begin{document}

\title{Intergalactic medium rotation measure of primordial magnetic fields}

\author[0000-0001-9786-8882]{Salome Mtchedlidze}
\affiliation{Dipartimento di Fisica e Astronomia, Universit\'{a} di Bologna, Via Gobetti 92/3, 40121, Bologna, Italy}
\affiliation{School of Natural Sciences and Medicine, Ilia State University, 3-5 Cholokashvili St., 0194 Tbilisi, Georgia}
\email{salome.mtchedlidze@unibo.it}

\author[0000-0001-7058-8418]{Paola Dom\'inguez-Fern\'andez}
\affiliation{Harvard-Smithsonian Center for Astrophysics, 60 Garden Street, Cambridge, MA 02138, USA}
\affiliation{Dipartimento di Fisica e Astronomia, Universit\'{a} di Bologna, Via Gobetti 92/3, 40121, Bologna, Italy}
\affiliation{INAF - Osservatorio di Astrofisica e Scienza dello Spazio di Bologna, via Gobetti 93/3, 40129 Bologna, Italy}

\author[0000-0003-0728-2533]{Xiaolong Du}
\affiliation{Department of Physics and Astronomy, University of California, Los Angeles, CA 90095, USA}

\author[0000-0002-3973-8403]{Ettore Carretti}
\affiliation{INAF - Istituto di Radioastronomia, Via Gobetti 101, 40129 Bologna, Italy}

\author[0000-0002-2821-7928]{Franco Vazza}
\affiliation{Dipartimento di Fisica e Astronomia, Universit\'{a} di Bologna, Via Gobetti 92/3, 40121, Bologna, Italy}
\affiliation{INAF - Istituto di Radioastronomia, Via Gobetti 101, 40129 Bologna, Italy}
\affiliation{Hamburger Sternwarte, University of Hamburg, Gojenbergsweg 112, 21029 Hamburg, Germany}

\author[0000-0002-3968-3051]{Shane Patrick O'Sullivan}
\affiliation{Departamento de Física de la Tierra y Astrofísica \& IPARCOS-UCM, Universidad Complutense de Madrid, 28040 Madrid, Spain}

\author[0000-0002-7304-021X]{Axel Brandenburg}
\affiliation{Nordita, KTH Royal Institute of Technology and Stockholm University, Hannes Alfv\'ens v\"ag 12, SE-10691 Stockholm, Sweden}
\affiliation{The Oskar Klein Centre, Department of Astronomy, Stockholm University, AlbaNova, SE-10691 Stockholm, Sweden}

\affiliation{McWilliams Center for Cosmology and Department of Physics, Carnegie Mellon University, 5000 Forbes Ave, Pittsburgh, PA 15213, USA}
\affiliation{School of Natural Sciences and Medicine, Ilia State University, 3-5 Cholokashvili St., 0194 Tbilisi, Georgia}

\author[0000-0003-0217-9852]{Tina Kahniashvili}
\affiliation{McWilliams Center for Cosmology and Department of Physics, Carnegie Mellon University, 5000 Forbes Ave, Pittsburgh, PA 15213, USA}
\affiliation{School of Natural Sciences and Medicine, Ilia State University, 3-5 Cholokashvili St., 0194 Tbilisi, Georgia}
\affiliation{E. Kharadze Georgian National Astrophysical Observatory, 0179, 47-57 Kostava St., Tbilisi,Georgia}

\begin{abstract}
The Faraday rotation effect, quantified by the rotation measure (RM), is a powerful probe of the large-scale magnetization of the Universe — tracing magnetic fields not only on galaxy and galaxy cluster scales but also in the intergalactic medium (IGM;  referred to as $\mathrm{RM}_{\text{IGM}}$).
The redshift dependence of the latter has extensively been explored with observations.
It has also been shown that this relation can help to distinguish between different large-scale magnetization scenarios.
We study the evolution of this $\mathrm{RM}_{\text{IGM}}$ for different primordial magnetogenesis scenarios to search for the imprints of primordial magnetic fields (PMFs; magnetic fields originating in the early Universe) on the redshift-dependence of $\mathrm{RM}_{\text{IGM}}$. We use cosmological magnetohydrodynamic (MHD) 
simulations for evolving PMFs during
large-scale 
structure formation, coupled with the light-cone analysis to produce a realistic statistical sample of mock $\mathrm{RM}_{\text{IGM}}$ images. 
We study the predicted behavior 
for the cosmic evolution of $\mathrm{RM}_{\text{IGM}}$ for different correlation lengths of PMFs, and provide fitting functions for their dependence on redshifts. 
We compare these mock RM trends with the recent analysis of the 
the LOw-Frequency ARray
(LOFAR) RM Grid
and find that large-scale-correlated PMFs should have (comoving) strengths $\lesssim 0.75 \nG$, 
if they originated during inflation with the scale-invariant spectrum and (comoving)  correlation length of $\sim 19 \cMpch$ or $ \lesssim 30 \nG$ if they originated during phase-transition epochs 
with the comoving correlation length $\sim 1 \cMpch$. 
Our findings agree with previous observations and confirm the results of semi-analytical studies, showing that  
upper limits on the PMF strength decrease as their coherence scales increase.
\end{abstract}


\section{Introduction} 
\label{sec:intro}
The ubiquitous presence of magnetic fields on small (such as, e.g., planets and stars) as well as on large scales (interstellar medium (ISM), galaxies, galaxy clusters; see, e.g., \citealt{GovFer2004,Kulsrudetal2007}, for reviews and references therein) raises interest in understanding the origin of the magnetic field. 
In magnetogenesis theories, a weak magnetic seed field is generated from a 
negligibly small 
initial field present 
either in the pre-recombination Universe 
during, e.g., 
inflation, reheating and preheating, 
phase transitions (cosmological, primordial scenarios) or later, during reionization and formation of the first structures (astrophysical scenarios; see \citealt{GrassoRubinstein2001,Kandusetal2011,Widrwoetal2012,DurrerNeronov2013,Subramanian2016,Vachaspati2020} for reviews).  
Subsequent amplification of this seed magnetic field is then expected during large-scale structure formation (see \citealt{Subramanian2019,BranNtormousi2023} for recent reviews).
Understanding the origin of the large-scale magnetization of the
Universe with at least $\sim \mathcal{O}(10)$ kiloparsec-correlated magnetic fields
of microgauss strengths found in galaxy clusters \citep{GovFer2004}, is a key science goal of the current \citep{Gaensleretal2010,Andersonetal2012,Lacyetal2020}
and upcoming radio surveys \citep[Square Kilometre Array (SKA), and its precursors and pathfinders][]{Hollittetal2015,Healdetal2020}.

High-energy gamma-rays from blazars, detected with the Fermi telescope (\citealt{Fermi_Lat2018}; capturing the low-end tail of the spectrum), High Energy Stereoscopic System \citep[H.E.S.S.;][capturing the high-end tail of the spectrum]{Abramowskietal_2014}, MAGIC \citep{Aleksicetal2010,Acciarietal2023}, VERITAS \citep{Fernandezetal2013,VERITAS2017}, and in the future, with the Cherenkov Telescope Array \citep[CTA,][]{Meyeretal2016,Abdallaetal2021}, are one of the powerful indirect probes of the strength and structure of extragalactic magnetic fields. It is still a subject of debate whether the deficit 
of the secondary emission (the low-end tail) in the blazar spectrum can be explained by plasma instabilities \citep{Broderick:2018nqf},
astrophysically generated seed fields, amplified and then transported to Mpc scales or by the volume-filling primordial magnetic fields \citep[PMFs;][]{Dolagetal2011}. 
In a recent combined analysis by the Fermi-LAT
and 
High Energy Stereoscopic System 
(H.E.S.S.) collaborations \citep{Aharonianetal_2023}, Mpc-correlated, volume-filling magnetic fields with a lower limit of $7.1 \times 10^{-16} \G$ or $3.9 \times 10^{-14} \G$ on their field strength have been favored, 
depending on the blazar activity periods (short or longer, respectively).\footnote{As clarified by \citealt{Aharonianetal_2023}, their constraints are not affected by plasma instability processes for the lower source variability period ($10 \yr$) of the blazars assumed in the analysis.}

The rotation measure (RM) of the rarefied, magnetized cosmic plasma
is another probe of the extragalactic magnetic field strength and structure
\citep{Sofueetal1968}. 
Generally, for a background source being at a cosmological distance
and viewed through a foreground magnetized plasma, the RM 
traces the
line of sight (LOS) magnetic field, $B_l$, where the LOS electron number
density, $n_\mathrm{e}$, is assumed to be known.
The value of RM is given by
\begin{equation}
\begin{aligned} 
\label{eq:RM} 
\mathrm{RM} &=\frac{e^3}{2\pi m_\text{e}^2c^4}\int_0^{l}  (1+z)^{-2}n_e(z)B_{l}(z)dl(z) &&\\\
&= 0.812 \int_{0}^{l} (1+z)^{-2} \bigg( \frac{n_e}{\text{cm}^{-3}} \bigg)\bigg( \frac{B_l}{\mu \text{G}} \bigg) \bigg( \frac{dl}{\text{pc}} \bigg) ~~\frac{\text{rad}}{{\text{m}}^{2}}, &&
\end{aligned}
\end{equation}
with $z$, $e$, $m_e$, and $c$  being the redshift, electron
charge, electron mass, and the speed of light, respectively.
RM quantifies\footnote{
Equation~(\ref{eq:RM}) is derived for the case when a wave propagates in the direction ($\hat{\mathbf{e_k}}$) parallel to the spatially homogeneous magnetic field, $\mathbf{B}$. However, it can be shown that this equation still holds for other situations \citep{RybickiLightman1979}  by replacing $B_l$ with $\mathbf{B} \cdot \mathbf{\hat{e}_k}$ \citep{Ferriereetal_2021}.
More complex analysis is required when computing the Cosmic Microwave Background (CMB) birefringence effect  \citep{KosowskyLoeb1996} by stochastic (statistically homogeneous and isotropic) PMFs: in the this case, RM depends not only on the amplitude of the magnetic field strength, $B_l$, but also on the magnetic field spatial distribution \citep[e.g., spectrum,][]{Campanelli:2004pm,Kosowsky:2004zh,Plancketal2016}; see also \cite{galli2021}, \cite{Mandal:2022tqu}, \cite{LiteBIRD:2024tmk} for the detection prospects of PMFs from such analysis.
} the change in the polarization angle, $\Delta \phi$,
\begin{equation}
\begin{aligned} 
\Delta\phi = \phi_{\text{intrinsic}} - \phi_{\text{measured}}  = \mathrm{RM} \lambda^2
\end{aligned}
\end{equation}
of the Faraday rotated polarized emission observed at a certain wavelength, $\lambda$ \citep{GardnerWhiteoak1963}.\footnote{In observations, observable polarization spectrum
$P(\lambda)^2 = \int_{\infty}^{\infty} F(\Phi)e^{2i\Phi\lambda^2} d\Phi$
is used to reconstruct the so-called Faraday dispersion spectrum $F(\Phi)$, where $\Phi$ itself defines the Faraday depth (i.e., the Faraday rotation at a certain distance along the LOS).  The Faraday depth is defined similarly to the RM (Equation~(\ref{eq:RM})), however, it coincides the RM only in the situation when a point-like source is viewed through a single, non-emitting, magnetized foreground medium; see, e.g., \cite{Ideguchietal_2018,Andersonetal2021}.}
While various analyses employing the Faraday rotation effect have been
used for reconstructing properties of the magnetic field, e.g., in the
intracluster medium \citep[see, e.g.,][]{Andersonetal2021},
constraining the value of the RM in the IGM
($\mathrm{RM}_{\text{IGM}}$), as well as (or in combination with) studying 
the $\mathrm{RM}_{\text{IGM}}$ evolution with redshift
has now become a promising alternative approach to search for large-scale
magnetic field imprints.
The total RM caused by the all the magnetized plasma along the LOS is
usually decomposed into the following contributions:
\begin{equation}
\label{eq:RM-contributions}
\mathrm{RM} = \mathrm{RM}_{\text{source}} + \mathrm{RM}_{\text{Gal}} + \mathrm{RM}_{\text{IGM}},
\end{equation}
where $\mathrm{RM}_{\text{source}}$  and $\mathrm{RM}_{\text{Gal}}$ 
are the contributions from the magnetized medium of the source itself and of our own Galaxy, respectively. 
The subtraction of $\mathrm{RM}_{\text{Gal}}$ from the total $\mathrm{RM}$ yields the so-called residual rotation measure (RRM), which accounts for the Faraday rotation effect caused by extragalactic magnetic fields. The RRM has extensively been studied through observations (see, e.g., \citealt{Fujimotoetal1971,Nelson1973a,Nelson1973b,Vallee1975,KronbergNormandin1976,Kronbergetal1977}
for some of the early works, and \citealt{AkahoriRyu2011,OSullivanetal2020,Carrettietal2022,Pomakovetal2022,Carrettietal2023}
for recent studies). 
As light propagates through a magnetized medium, its polarization angle undergoes Faraday rotation; since we expect the magnetic field orientation to vary along the LOS --- resulting in RMs having both negative and positive values --- the average RRM should be close to zero.
However, a redshift dependence of the higher-order statistics of RRM, 
such as its variance and kurtosis,
is expected in the presence of IGM magnetic fields \citep{Nelson1973a}. Thus, since the
light emitted from high-redshift sources spends more time in the foreground magnetized medium than the light from the low-redshift sources, the variance of the RRM for high-redshift sources
is anticipated to be larger compared to the variance of the RRM for low-redshift sources \citep{Nelson1973a,AkahoriRyu2010,AkahoriRyu2011}.

The analysis of the emission from various extragalactic sources (such as,
e.g., quasars and fast radio bursts) has yielded controversial
results over whether the variance of RRM evolves with redshift, and if it
does, whether its evolution is solely attributable to the IGM component
\citep{Nelson1973a,Akahorietal2016}.
Some authors \citep{Reinhardt1972,Vallee1975,KronbergNormandin1976,Kronbergetal1977,Orenetal1995,Hammondetal_2012,Carrettietal2022,Carrettietal2023,Manningsetal2023} have found no clear evidence of an increase of the RRM with redshift while placing constraints on its contribution to the total RM \citep{Vallee1975,Kronbergetal1977} to be less than $<10\,\textrm{rad m}^{-2}$.
A similar constraint was obtained in \cite{OSullivanetal2020} where a statistical difference in the RM between physical (extragalactic sources being at the same redshift) and random (sources at different redshifts) pairs was used to isolate the $\mathrm{RM}_{\text{IGM}}$ contribution
\citep[for pioneering work, see also][]{Vernstrometal2019};
since it is expected that an RM difference between physical pairs should be smaller compared to the RM difference for sources located at different redshifts, then the difference $(\Delta \mathrm{RM}_{\mathrm{rms}}^{\mathrm{random}} - \Delta\mathrm{RM}_{\mathrm{rms}}^{\mathrm{physical}} )$ is regarded as an 
RM induced by the IGM magnetic field. Conversely, other authors claimed the RRM dependence on redshift
\citep{Welteretal1984,Sofueetal1968,Kawabataetal1969,ReinhardtThiel1970,Fujimotoetal1971,ThomsonNelson1982, Kronberg2008,Neronovetal2013,XuHan2014,Pshirkovetal2016,Pomakovetal2022},
with its value saturating at $z \gtrsim 1$ \citep{XuHan2014,Pshirkovetal2016}.  A semi-analytical approach of \cite{Kronbergetal1977}, \cite{Pshirkovetal2016} as well as the light-cone analysis within cosmological simulations \citep{AkahoriRyu2011} confirmed the evolution of $\mathrm{RM}_{\text{IGM}}$ with redshift. In \cite{Pshirkovetal2016} and \cite{Carrettietal2023}, it was further shown that the $\mathrm{RM}_{\text{IGM}}$ evolution trends depend on the structure of the intervening magnetic field, and therefore, on the IGM magnetization scenarios \citep{Vazzaetal2017,Locatellietal2018,Garciaetal2023,Carrettietal2023}.

Motivated by the results of the aforementioned studies and the recent
findings of \cite{Carrettietal2022}, which favor the IGM over the source
contribution in the RRM (observations in the 144 MHz regime) and hint at
the existence of the ordered, large-scale magnetic fields in filaments
\citep[similar to outcomes of][]{Vernstrometal2021}, we
explore in this paper the evolution trends of $\mathrm{RM}_{\text{IGM}}$ for different
primordial magnetogenesis scenarios.
We study similar PMF models and a cosmological simulation setup as in \cite{Mtchedlidzeetal_2022} (hereafter, Paper~I) along with a light-cone analysis \citep{yt-Turk2011} to generate mock $\mathrm{RM}_{\text{IGM}}$ maps for $z\leq 2$ redshift depths.

The structure of the paper is as follows: in Section~\ref{sec:methods}
we describe our simulation setup, PMF models, and the light-cone generation
technique; in Sections~\ref{sec:Results} and \ref{sec:NumAsp}
we present our results and discuss simulation and analysis uncertainty
aspects, respectively, and in Section~\ref{sec:Summ} we summarize our work.

\section{Methods}
\label{sec:methods}

\subsection{Simulation setup and initial conditions}
\label{subsec:SetupInits}
\begin{deluxetable*}{c| c | c c c c c c c}[ht]
\tablecaption{Initial conditions for the magnetic field. 
The correlation length and the mean value of the smoothed
(on a $1 \cMpch$ scale) magnetic field are denoted
by $\lambda_{\rm B}$ and $B_{1 \rm Mpc}$, respectively, while
$\langle B_{0}^2 \rangle$ and $\langle B_{0} \rangle$ are the
means of the initial magnetic field energy and the initial
magnetic field strength, respectively. 
All characteristics derived here use the comoving magnetic field strength. 
\label{tab:Tab1}
}
\tablehead{
\colhead{Scenario} & \colhead{Model} &  \colhead{Simulation ID}& \colhead{Normalization} & \colhead{$\langle B_{0}^2 \rangle$}  & \colhead{$\langle B_{0} \rangle$} &  \colhead{$B_{1 \rm Mpc}$}&\colhead{$\lambda_{B}$} \\
\colhead{} &\colhead{}&\colhead{~} &\colhead{[$\nG$]}  & \colhead{ [(nG)$^2$]}  &\colhead{[nG]} &\colhead{[nG]} & \colhead{[$h^{-1}\cMpc$]} 
}
\startdata
\multirow{4}{*}{Inflationary}  & \multirow{1}{*}{(i) Uniform}             &   u   & 1 & 0.72                     & 0.85   & 0.85      &---\\
                              \cline{2-8}
                             &  \multirow{4}{*}{(ii) Scale invariant}   &      & 0.01 &  0.72$\times 10^{-5}$    & 0.78$\times 10^{-3}$   & 0.0078   & 18.80\\
                              &                                         & km1 & 0.1   & 7.2$\times 10^{-3}$                  & 0.0078    & 0.0079    & 18.80\\
                              &                                         &     &  1    & 0.72                    & 0.78     & 0.79     & 18.80\\
                               &                                         &     &  50    & 1.8 $\times 10^{3}$    & 39.1     & 39.6     & 18.80\\
\hline
\multirow{11}{*}{ Phase transitional}   &  \multirow{4}{*}{(iii) $\lambda_{\text{peak} } = 4.92 \cMpch$ } &                    & 1   & 0.72    & 0.78   & 0.79    & 3.49  \\
                                     &                                                                    &  k25                & 5    & 17.9   & 3.91    & 3.93    & 3.49\\
                                     &                                                                    &                     & 10   & 71.87   & 7.81    & 7.85     & 3.49\\
                                     &                                                                    &                    &  15   & 161.7  & 11.71   & 11.78   & 3.49\\
                                    \cline{2-8}
                                    & \multirow{2}{*}{(iv) $\lambda_{\text{peak} } = 2.53 \cMpch$ }     &  k50                  & 0.01  &0.72$\times 10^{-4}$  & 0.0078 & 0.0078 & 1.81\\
                                    &                                                                  &                        &    1  & 0.72  & 0.78    & 0.78 & 1.81\\
                                    \cline{2-8}
                                    & \multirow{5}{*}{(v) $\lambda_{\text{peak} } = 1.26 \cMpch$ }  &                        & 1    & 0.72   & 0.78 & 0.78 & 1.00\\
                                    &                                                                &                        & 5    & 17.9   & 3.91    & 3.91   & 1.00\\
                                    &                                                                  &  k102                 & 10   & 72  & 7.81   & 7.8   & 1.00\\
                                    &                                                                 &                        & 15   & 1.62 $\times 10^2$ & 11.7    & 11.7   & 1.00\\
                                    &                                                                 &                        & 50   & 1.8 $\times 10^{3}$ & 39.1  & 39.1  & 1.00\\
\enddata
\end{deluxetable*}
We use the magnetohydrodynamic (MHD) cosmological code \texttt{Enzo} \citep{Bryanetal2014} to simulate a $(135.4 h^{-1}\cMpc)^3$ (``c'' referring to comoving units) comoving volume employing $1024^3$ grid points and $1024^3$ dark matter (DM) particles 
with a $132\,\ckpch$ and $m_{\text{DM}} = 2.53\times 10^{8} M_{\odot}$ spatial and DM mass resolutions, respectively. We assume the Lambda cold dark matter ($\Lambda$CDM) cosmology with the parameters $h=0.674$,
$\Omega_m=0.315$, $\Omega_b=0.0493$, $\Omega_{\Lambda}=0.685$, and $\sigma_8=0.807$ \citep{Planck2018}.
In this work, we double the simulated volume (compared to our previous setup in Paper I) to produce deep light cones (see below). Nevertheless, the temporal \citep[second-order Runge-Kutta scheme, RK,][]{SHU1988439} and the spatial \citep[piecewise linear method, PLM,][]{1979JCoPh..32..101V,COLELLA1985264} reconstruction schemes, as well as the Riemann solver \citep[Harten–Lax–van Leer, HLL,][]{Toro1997} are the same as in Paper~I.
Similarly to our previous work, we use the Dedner cleaning algorithm \citep[][]{Dedneretal2002} to keep the divergence of the magnetic field at its minimum and focus on the ideal, adiabatic physics.

The magnetic field models that we aim to constrain are as follows:
\begin{enumerate}
	\item[1.] Uniform (constant-strength) field that corresponds to the inflationary magnetogenesis according to the Mukohyama model \citep{Mukohyama2016}.
	\item[2.] The nonhelical, scale-invariant field which is predicted by some of the inflationary magnetogenesis models \citep[see, e.g.,][for the Ratra and Ratra-like models\footnote{These models
require breaking of the conformal invariance of the electromagnetic action in order for weak seed magnetic fields to be amplified
during the accelerated expansion phase of the Universe.}]{Ratra1992,Kanoetal2009,Emamietal2010,TomohiroMukohyama2012}. In this case, the field is stochastic, statistically homogeneous and is characterized by a scale-invariant ($\sim k^{-1}$) spectrum.\footnote{We note that,
unlike Paper I, here we use an initial magnetic field with a truly
scale-invariant spectrum and random phases rather than one that was only
originally scale-invariant, which later developed a nearly Kolmogorov-like
spectrum with non-random phases.}
	\item[3.] Nonhelical field with a characteristic peak in the energy spectrum. Depending on the peak scale (or the correlation length), these models can be motivated by either inflationary or phase-transitional magnetogenesis. We explore the latter, assuming magnetic correlation lengths of $3.49, 1.81$, and $1.00 \cMpch$. 
 However, we note that in general,
    there is no clear bound between the correlation lengths of inflation- and phase transition-generated magnetic fields in the literature. Nevertheless, in this study, we associate models with a characteristic peak to the phase-transitional scenario to distinguish them from the larger-scale-correlated magnetic fields.
\end{enumerate}

\begin{figure}[t]
    \centering
    \includegraphics[width=7.6cm]{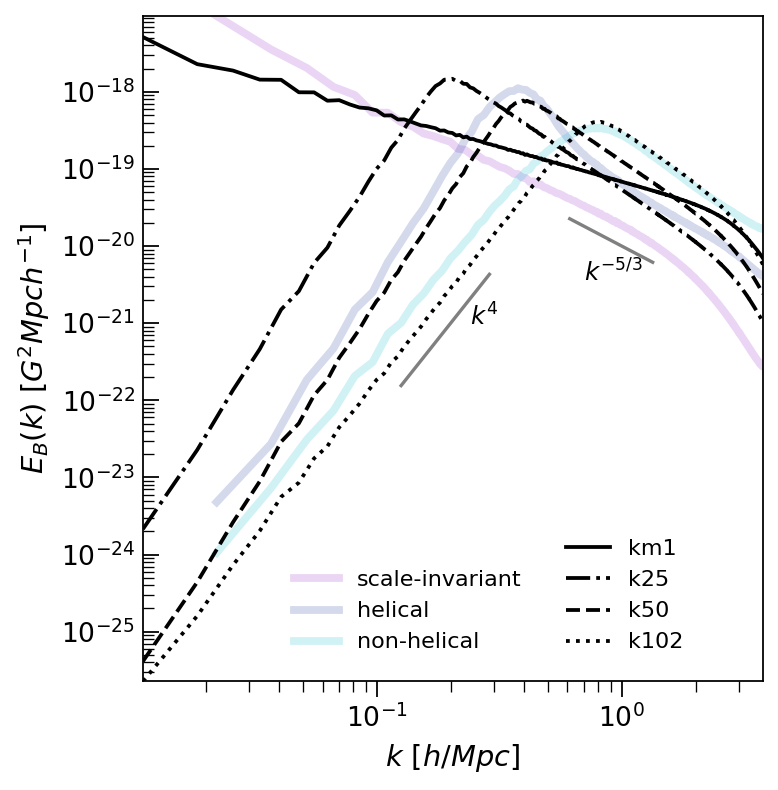}
   \caption{The initial ($z=50$) magnetic power spectra for the stochastic setups (black lines) with respect to the initial conditions used in Paper I (low-opacity lines). 
   }
    \label{fig:B-PS_inits}
\end{figure}

We generate magnetic field distributions for the latter two cases
using the \textsc{Pencil Code} \citep{JOSS} initialization routine and show their spectra in
Figure~\ref{fig:B-PS_inits}.
The shapes of the power spectra for the k50 and k102 models are the
same as the shapes of the power spectra in the helical and nonhelical cases
(Paper I), respectively. However, in this work, all of our PMF models
are initialized as Gaussian random fields (contrary to the initial conditions of Paper I).
As in paper~I, we do not account for PMF-induced
perturbations on the matter power spectrum.
\cite{Wasserman1978}, \cite{Kimetal1996}, \cite{Kahniashvilietal2013}, \cite{Sanatietal2020}, \cite{Katzetal2021}, and \cite{Pavicevicetal_2024}
have studied such perturbations (sourced by the Lorentz force), demonstrating that they produce additional clustering of matter on dwarf galaxy scales
\citep[$\sim 10 \kpc$,][]{Sanatietal2020,Katzetal2021}. 
These scales are much smaller than the adopted resolution in this work.

In Table~\ref{tab:Tab1} we provide a list of the initial characteristics of the studied scenarios.
The normalization is such that the mean magnetic energy 
and field strength (averaged over the whole simulated volume) are the same for all models. 
Throughout this study, we mainly use simulations with a normalization of $1 \nG$ (referred to as the main run), unless otherwise specified. In this case, the mean magnetic field strength, being similar to the smoothed amplitude of the field on $1 \Mpc$ scales, is below the constraints derived from the cosmic microwave background (CMB) analysis; although it is worth noting that in this analysis, upper limits on the smoothed value of the field are obtained for PMFs with a simple power-law spectrum \citep{Planck2018},\footnote{PMFs might also be constrained through their impacts on recombination \citep{Lynch:2024} and (re)ionization \citep{Paolettti:2022}, and serve to relax Hubble tension \citep{JedamzikPogosian2020}. The amplitude of our PMFs is higher than the upper limits derived from accounting for PMF-induced baryon clumping across the recombination epoch \citep{Jedamzik:2018itu}.}
while our small-scale stochastic models feature more complex power spectra.

\begin{figure*}[t]
    \centering
    \includegraphics[width=18.5cm]{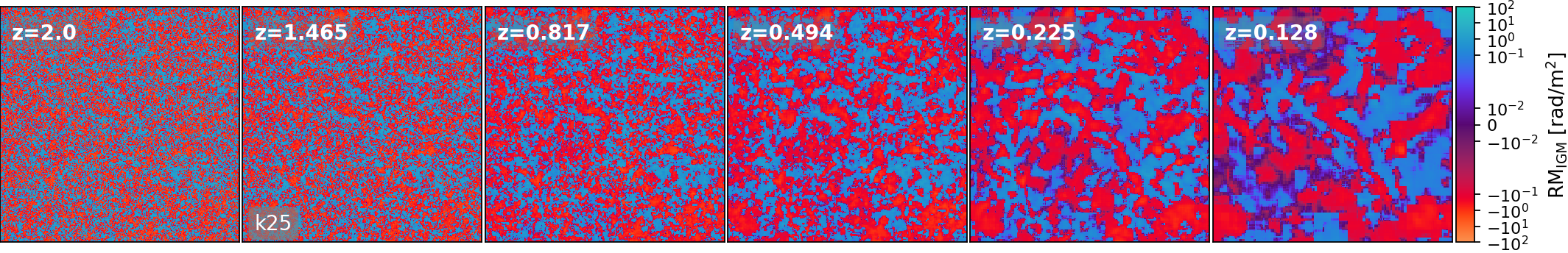}
    \vspace{-5.4pt}
    \includegraphics[width=18.5cm]{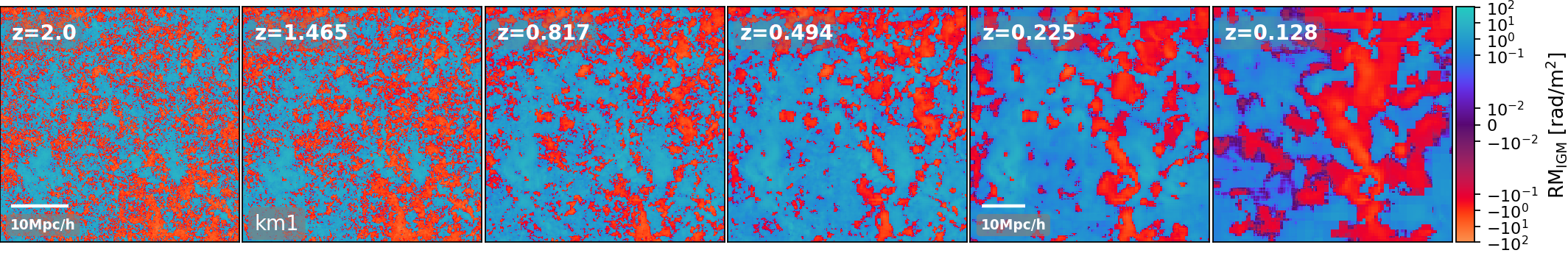}
    \vspace{-1.0pt}
    \includegraphics[width=18.5cm]{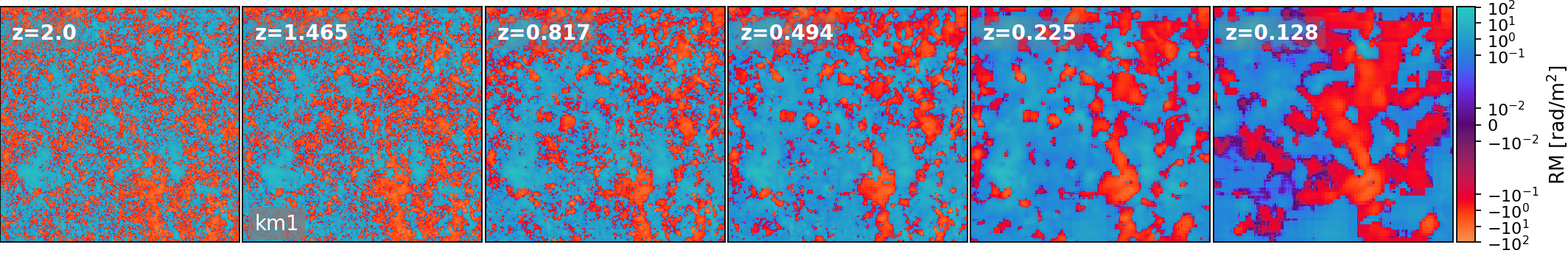}
    \includegraphics[width=18.5cm]{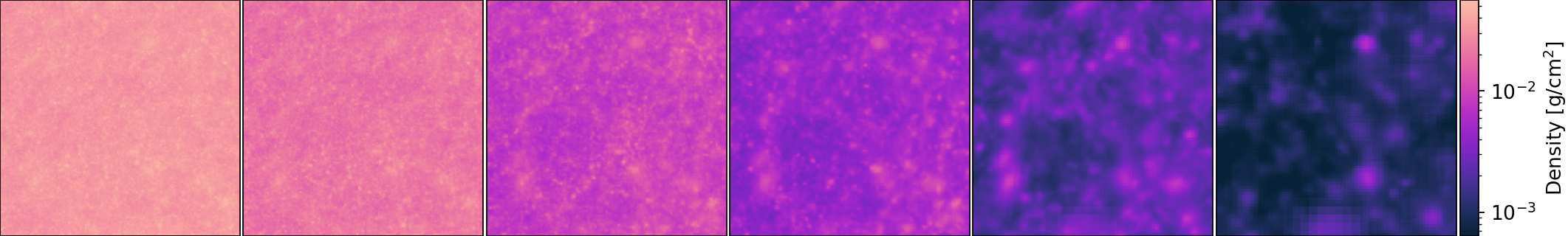}
   \caption{RM (first three rows from the top) and density (last row) maps for different redshift depths as are seen by an observer at $z=0$ with a \SI{2}{\degree} FOV. The RM maps for the regions excluding galaxy clusters (referred to as $\mathrm{RM}_{\text{IGM}}$) are shown in the first and second rows for the k25 and scale-invariant models, respectively; the third row shows the RM for the whole LOS for the scale-invariant case.}
    \label{fig:projs}
\end{figure*}

\subsection{Light cones}
\label{subsec:LightCones}

We use a modified version of \texttt{yt\_astro\_analysis}, an extension \citep{Smithetal_2022} of the \textit{yt} toolkit \citep{yt-Turk2011}, for producing a stacked sequence of the simulated boxes to integrate Equation~(\ref{eq:RM}) from an observer redshift $z_\text{obs}=0$ to redshift $z_\text{far}$ = 2. 
The purpose of using this technique is to ameliorate our previous constraints (Paper I) on PMFs by using a more accurate approach.
We proceed with the following steps: (i) 
simulation boxes are stacked based on the output redshifts,
(ii) Equation~(\ref{eq:RM}) is then integrated for each dataset in the stack, and finally, (iii) light-cone images are produced  when the desired field of view (FOV)
and the resolution of the image is specified.
The advantage of the \textit{yt} method is that it minimizes the likelihood that the same cosmic-web structures are sampled more than once along the LOS. This is ensured by varying the projection axis and the center of the projected region; see Section~7.6 of \cite{yt-Turk2011} for more details on the method. Such a method further allows us to produce different light-cone realizations for each PMF model studied in this work.
The redshift interval chosen in our analysis matches the redshift span of the LOFAR Two-metre
Sky Survey (LoTSS) survey analysis \citep{Carrettietal2023}.

We use a total of 54 redshift snapshots for our stacking procedure while providing the 
\SI{2}{\degree} FOV and \SI{20}{\arcsecond}
image resolution, which is similar to the resolution of the LoTSS Data Release 2 \citep[DR2;][]{Shimwelletal2019,Shimwelletal2022,OSullivanetal2023}, and is larger than the spatial resolution of our simulations at $z\gtrsim 0.6$. 
The provided FOV is always smaller than the proper width of the simulated volume at all redshifts, and it determines the fraction of the box width used for producing RM images. 
We produce RM images for low-density regions, referred to as $\mathrm{RM}_{\text{IGM}}$ and satisfying the $\rho / \langle \rho \rangle < 1.3\times 10^{2}$ criteria, with $\rho$ being the density field and also for the whole LOS, without filtering any regions. We chose the $\rho / \langle \rho \rangle < 1.3\times 10^{2}$ criterion to exclude massive objects from our $\mathrm{RM}_{\text{IGM}}$ analysis while still including the warm-hot ionized medium (WHIM). The same approach has been used in observation analysis \citep{Carrettietal2022}.

Generating $\mathrm{RM}$ images using the light-cone approach has previously been done in studies such as those of \cite{AkahoriRyu2011,OSullivanetal2020,Vazzaetal2020}. \citealt{AkahoriRyu2011} used hydrodynamical cosmological simulations where magnetic fields were estimated from the (turbulent) kinetic energy; this was motivated by \citealt{Ryuetal2008}, who showed that turbulent flows, generated due to cosmological shocks during structure formation amplify weak seed fields and thus trace magnetic fields. A recent advancement in this method has been made by \citealt{OSullivanetal2020} and \citealt{Vazzaetal2020}. The authors of these works employed MHD simulations along with light-cone analysis to generate mock RM maps. Our method presents an improvement of the techniques of \cite{AkahoriRyu2011}, \cite{OSullivanetal2020}, and \cite{Vazzaetal2020}. 
In particular, unlike these previous studies,
we output $\mathrm{RM}_{\text{IGM}}$ maps
throughout the redshift span $z_\text{far}$ to $z_\text{obs}$ without the need for replication of the simulated data.
Additionally, we provide a statistical sample of the RM sky through light-cone realizations.

\section{Results}
\label{sec:Results}

\subsection{RM evolution}
\label{subsec:RmrmsEv}

In Figure~\ref{fig:projs} we show the simulated RM images produced
with a $\SI{2}{\degree}$ FOV along with density maps for different redshift depths. Figure~\ref{fig:projs} illustrates differences between the $\mathrm{RM}_{\text{IGM}}$ maps of small- and large-scale-correlated PMFs. First, we see that RM maps of the k25 model (with an initial $\sim 3.5 h^{-1}\cMpc$ correlation length) compared to the RM maps of the km1 case (with an initial $\sim 19 h^{-1}\cMpc$ correlation length) show smaller correlated structures. This feature is more pronounced at higher redshift depths due to the larger difference in magnetic correlation lengths between these models at early times. Therefore, high-redshift RM data will be more suitable for distinguishing different PMF models. The 
sizes of the correlated structures in the $\mathrm{RM}_{\text{IGM}}$ and total $\mathrm{RM}$ maps are similar.

\begin{figure}[htbp]
    \centering
    \includegraphics[width=8.3cm]{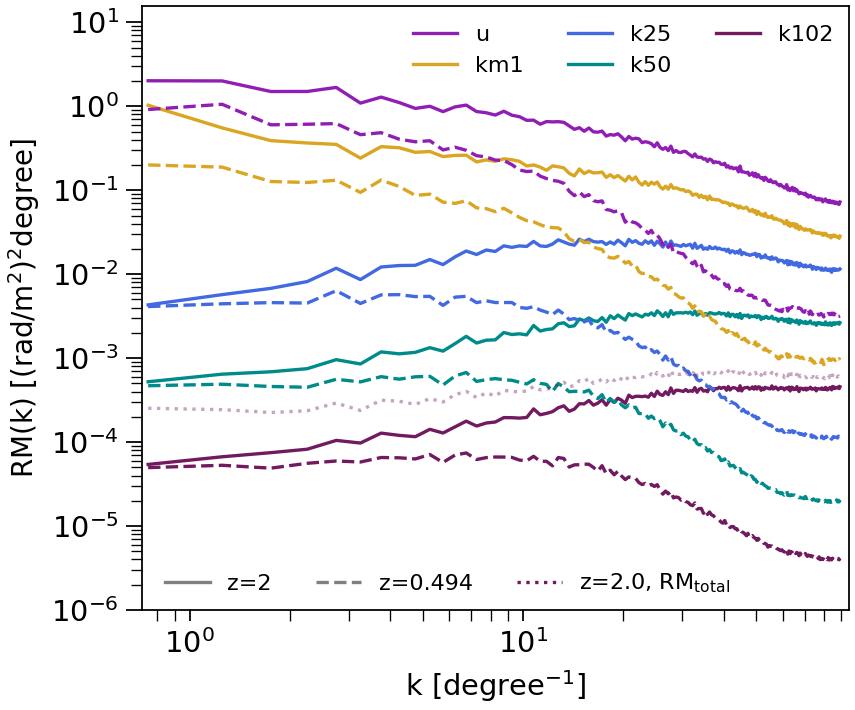}
    \caption{RM (two-dimensional) power spectra for different PMF models at $z=2$ (solid) and $z=0.494$ (dashed) redshift depths. For the k102 model, we also show the total (without excluding any regions) RM power spectra integrated up to $z=2.0$ redshift depth (dotted lines). 
    } 
    \label{fig:RM-PS}
\end{figure}

Figure~\ref{fig:RM-PS} provides a quantitative illustration of the differences between the $\mathrm{RM}_{\text{IGM}}$ maps for different PMF models. As shown, both the shape and the amplitude of the $\mathrm{RM}_{\text{IGM}}$ power spectrum (averaged over different light-cone realizations) are distinguishable for inflationary and phase-transitional models. The uniform and scale-invariant models have the largest power on all angular scales compared to the k25, k50, and k102 models whose coherence scales are smaller. We also observe that small-scale PMFs (most of the magnetic energy concentrated on small scales; see Figure~\ref{fig:B-PS_inits} and Figure~\ref{fig:PS-redshift} in Appendix~\ref{app:Bfield-PS}) show more power on small scales at $z=2$ redshift depths. 
As redshift depth increases, power is added on smaller scales for all models because high-redshift structures have smaller angular sizes.
At lower redshift depth ($z=0.494$), the amplitude of $\mathrm{RM}_{\text{IGM}}(k)$ decreases for all models, while the small-scale PMFs show a peak at $\sim \SI{0.17}{\degree}$ (k102 model) corresponding to $\sim 3.6 \Mpc$ ($2.43 \Mpch$) at that redshift. We also checked peak scales, $1/k_{\text{peak}}$, at $z=1.007$ where the differences between the peak scales of k25, k50, and k102 models are more pronounced. For k25 we obtain the largest peak scale ($1.665 \Mpch$) (see also \citealt{AkahoriRyu2011}, who argued that the peak of the RM power spectrum reflects the scale of filaments). 
Finally, we also note that a comparison between the $\mathrm{RM}_{\text{IGM}}$ power spectrum and the total RM power spectrum ($\mathrm{RM}_{\text{total}}$) shows only minor differences in shapes (at large angular scales) for small-scale models, while for large-scale fields, the differences are mainly seen in amplitude (we show $\mathrm{RM}_{\text{total}}$ only for the k102 model in the figure).

\begin{figure}[t]
    \centering
    \includegraphics[width=8.5cm]{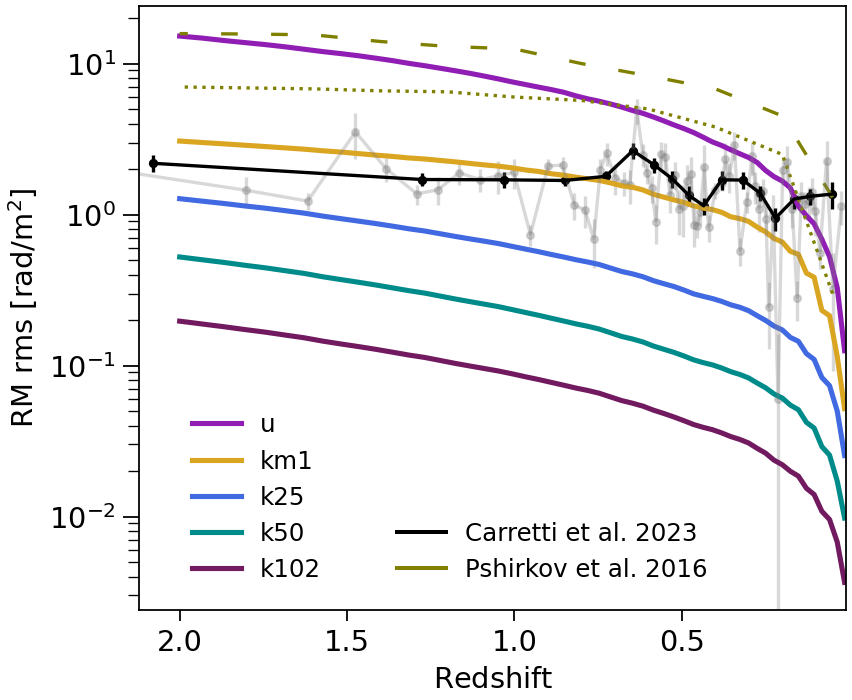}
    \caption{Redshift evolution of the $\mathrm{RM}_{\text{IGM}}$ of different PMF models (our main runs, $1\nG$ normalization). The black and gray lines show observed trends of RM rms from \cite{Carrettietal2023} when using 60 and 15 sources per redshift bin (the LoTSS data), respectively. The light green-dashed and dotted lines show theoretical estimations for Hubble- ($\sim 4200 \Mpc$) and Jeans-scale ($2.3 \Mpc$) magnetic fields from \cite{Pshirkovetal2016}. In this latter work, magnetic fields were calculated from the density field, the redshift dependence of which was drawn from a lognormal distribution. 
    } 
    \label{fig:RM-rmsEv-all}
\end{figure}

In Figure~\ref{fig:RM-rmsEv-all}, we show the evolution of the $\mathrm{RM}_{\text{IGM}}$ rms statistics for all PMF models along with the LoTSS data from \cite{Carrettietal2023} and theoretical estimations of RM from \cite{Pshirkovetal2016}. 
We calculated RM-rms values for each redshift depth from the corresponding $\mathrm{RM}_{\text{IGM}}$ images (see Figure~\ref{fig:projs}).
Each RM-rms line in Figure~\ref{fig:RM-rmsEv-all} 
corresponds to the average RM rms of 10 light-cone realizations; 
the probability density functions (PDFs), obtained from these different realizations are similar and their averages are not FOV dependent. At high-redshift depths, the PDFs for stochastic models show better fits with the Gaussian function, while the uniform model is better fitted by a lognormal distribution
(lognormal fit of the $|\textrm{RM}|$ PDF has also been found in \cite{AkahoriRyu2010} and \cite{AkahoriRyu2011}).
We refer the reader to Appendices \ref{app:PDFs-FOV-seeds} and \ref{app:PDFs-evolution}, where we show the outcomes of the aforementioned analysis (see Figures~\ref{fig:RM-PDF-FOVdep} and \ref{fig:RM-pdfEv}). 
We also note that the LoTSS data shown in Figure~\ref{fig:RM-rmsEv-all} has been corrected for the $A_{\text{rrm}} (1+z)^{-2}$ term, with $A_{\text{rrm}} = 0.6$.  We subtracted this term from the observed RM \citep{Carrettietal2023} in order to focus on the $\mathrm{RM}_{\text{IGM}}$ contribution.
This correction allowed us to compare the observed and simulated RM evolutions; see Equation (18) in \cite{Carrettietal2023} and the corresponding discussion for more details.

As can be seen from Figure~\ref{fig:RM-rmsEv-all}, the evolution of $\mathrm{RM}_{\text{IGM}}$ rms values
is distinguishable for small- (k25, k50, k102) and large-scale-correlated (u, km1) magnetic fields. 
The RM rms value increases with redshift in all our PMF models, 
although the uniform model shows the highest RM values and 
faster growth compared to the growth of the km1 model. 
We also see that 
RM trends shift toward higher rms values when considering stochastic models with larger coherence scales. This is in agreement with our previous results (Paper~I), showing that larger-scale magnetic fields are prone to more efficient growth in filamentary structures and consequently, they lead to the largest RMs in the WHIM (see Figure~11 of Paper~I). 
The differences observed in the $\mathrm{RM}_{\text{IGM}}$ growth rates between our large-scale uniform and scale-invariant models can be attributed to the presence of less coherent structures in the LOS magnetic field for the latter model (see, e.g., Figure 9 of Paper I).
This argument also explains faster RM growth of the large-scale field (with coherence scales of the order of Hubble scale) with respect to the growth of smaller-scale fields (with coherence scales of the order of Jeans scale) 
found in \cite{Pshirkovetal2016}, and similarity for our (uniform model) and their (Hubble-scale field) results at $z=2$. 
We also note that the increase of RM-rms with increasing redshift depth is mainly a result of random-walk processes when integrating RM along LOS, see e.g., \cite{Nelson1973a}, \cite{AkahoriRyu2010}. As the number of structures along the line of sight decreases at high redshifts,  RM-rms is expected to saturate \citep{AkahoriRyu2011}. However, we find in our analysis that at $z = 2$, our simulated WHIM is still dense enough to induce RM-rms variations.

Similar differences between the RM trends of primordial uniform and smaller-scale-correlated models have been found in studies by \cite{Garciaetal2023} and \cite{Carrettietal2023}.
These authors have shown that stochastic models (in their case characterized by a simple power-law spectrum)
or astrophysical sources of magnetization \citep{Garciaetal2023,Carrettietal2023} lead to lower RMs compared
to the RMs from the uniform model, although in their case, the uniform model also shows the fastest growth.

\begin{deluxetable}{c c c c}
\tablecaption{
\label{tab:Tab2}
Fitted values for $\mathrm{RM}_{\text{IGM}}(1+z)$ dependence when using $f_{\text{fitting}}= A [\mathrm{log}_{10}(1+z)]^{\alpha}(1+z)^{\beta}$ fitting function; see also Figure~\ref{fig:RM-Ev-fittedLogx}. 
}
\tablehead{
\colhead{model} & \colhead{A} &  \colhead{$\alpha$} & \colhead{$\beta$}}
\startdata
uniform          &  9.450      & 0.986 & 0.876 \\
km1              &  5.234      & 0.861 & 0.070 \\
k25              &  0.211      & 0.603 & 1.624  \\
k50              &  0.058      & 0.586 & 1.908 \\
k102             &  0.023      & 0.602 & 1.883 \\ 
\enddata
\end{deluxetable}
RM-rms trends are well fitted by a logarithmic function:
\begin{equation}
\label{eq:fittingFun} 
f_{\text{fitting}}= A  \big[ \rm{log}_{10}(1+z)^{\alpha} \big] (1+z)^{\beta},
\end{equation}
as it is shown in Figure~\ref{fig:RM-Ev-fittedLogx} of Appendix~\ref{app:PDFs-evolution}. The corresponding fitted parameters — $A$, related to normalization, $\alpha$, and $\beta$ — for each PMF model are listed in Table~\ref{tab:Tab2}. The fitted $\alpha$ parameter determines the $\mathrm{RM}_{\text{IGM}}$ declining trend at low ($z\lesssim 0.75$) redshifts; higher values of $\alpha$ correspond 
to a faster decrease of $\mathrm{RM}_{\text{IGM}}$ at later times. 
The fitted parameter $\beta$ can be interpreted as a 
declining rate showing the fastest decline rates (the largest values of $\beta$) for small-scale stochastic models (k25, k50, k102). Although the $\beta$ values for the k50 and k102 models are hardly distinguishable, differences between the slopes of small- and large-scale correlated models are more pronounced; this is in agreement with the outcomes of our previous work (see Table 3 in Paper I). 
We also observe that the scale-invariant case exhibits the smallest declining rate, characterized by the smallest $\beta$.
This feature is already evident in Figure~\ref{fig:RM-rmsEv-all}, where the RM-rms trend of the scale-invariant model flattens at high redshifts.
Finally, we note that while the trend for the uniform model agrees with the simulation results of \cite{Garciaetal2023} and \cite{Carrettietal2023} (not shown in the figure), 
the results of our k50 model (green solid line) are at odds with those of \cite{Pshirkovetal2016} (green-dotted line). 
This inconsistency arises despite the similarity in coherence scale between our k50 model ($\lambda_{\rm B} \sim 2 \cMpch$) and the one with $\lambda_{\rm B} \sim 2.3 \Mpc$ studied in \cite{Pshirkovetal2016}
and is likely attributed to the $B\sim \rho^{2/3}$ scaling employed in the latter work. This scaling results from the assumption of isolated, isotropic spherical collapse of gravitating regions (conserving its mass and magnetic flux). In Paper I, we showed that small-scale models are not characterized by $B\sim \rho^{2/3}$ scaling either in collapsed regions or in filaments (their slopes are shallower; see also the discussion in Section~\ref{sec:NumAsp}). It has also been shown that the slope of $B-\rho$ relation depends on both magnetic field orientation and the geometry of collapsing regions \citep[see][and references therein]{Tritsisetal2015}.

The most important outcome of Figure~\ref{fig:RM-rmsEv-all} is that the $\mathrm{RM}_{\text{IGM}}$ values, extracted from the LoTSS RM catalog \citep{Carrettietal2023}, rule out 
the uniform model with an initial $1 \nG$ normalization 
while the shape of the RM-rms trend in the scale-invariant case matches the observation results best at high redshifts. 
The uniform model, although with a $0.1 \nG$ normalization, has also been excluded by \cite{Carrettietal2023}.
The authors of this work used 60 source/bin analysis for constraining their magnetogenesis models. In the rest of our discussion, we will also focus on the 60 source data. The purpose of this choice is not only to ease the comparison of the LoTSS data with all of our PMF model trends, but also a preference for the comparison of high redshifts ($z \gtrsim 0.5$).
The error in the RM data, as well as our analysis at lower redshifts, can be affected by environmental selection effects. Therefore, we choose the 60 source data from \cite{Carrettietal2023} since it has a lower error at high redshifts; besides this, this trend is not significantly different from the trend of 15 source/bin data at these redshifts.
\begin{figure}[htbp]
    \centering
    \includegraphics[width=8.5cm]{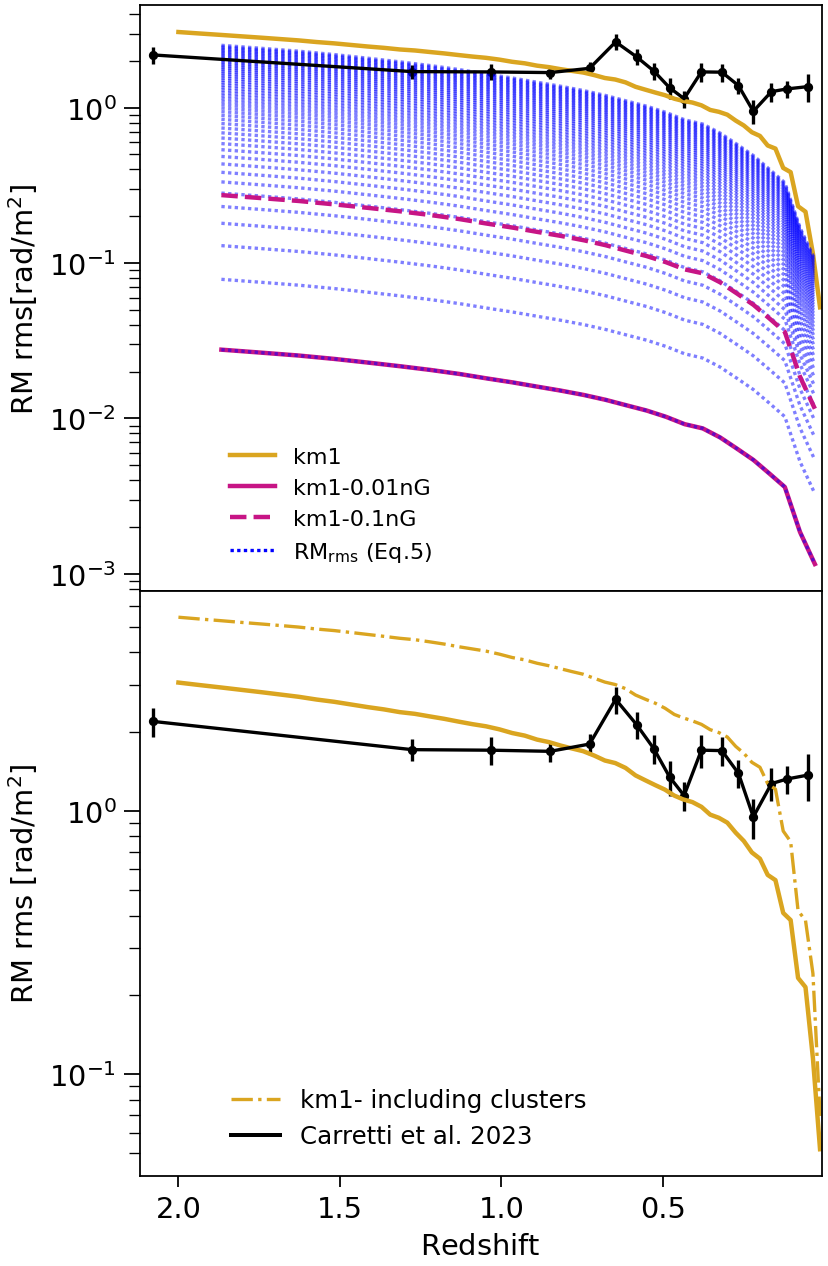}
    \caption{Redshift evolution of the $\mathrm{RM}_{\text{IGM}}$ of the scale-invariant (km1) PMF model. In the bottom panel, we show the RM-rms trends (dashed-dotted lines) for the whole LOS; in the top panel, we show RM-rms trends for different initial normalizations of the magnetic field,  $0.01 \nG$ (pink solid) and  $0.1 \nG$ (pink dashed lines), and those using Equation~(\ref{eq:RM-growth}) (densely dotted lines) for different ranges of magnetic field strength ($< 1 \nG$).}  
    \label{fig:RM-rmsEv-km1}
\end{figure}

We expanded our analysis by considering different initial normalizations for our PMF models.
We ran additional simulations with different initial amplitudes of the magnetic field (see Table~\ref{tab:Tab1}) to check how the shape and amplitude of RM-rms trends are influenced by the magnetic field strength. 
Considering that for certain initial normalizations of the magnetic field, the mean field strength is too low to significantly affect gas dynamics (see also below), we can assume that 
$\rm{RM}_{\rm {rms}}$ remains similar to the shape of the RM-rms for the $1\nG$ normalization,
$\rm{RM}_{\rm {rms, 1 nG}}$; then we can express this relationship as
\begin{equation}
\label{eq:RM-growth} 
\rm{RM}_{\rm {rms}} = \rm{RM}_{\rm {growth}}\bigg( \frac{B}{1\nG} \bigg) \, \rm{RM}_{\rm {rms, 1 nG}},
\end{equation}
where $\rm{RM}_{\rm {growth}}$ is the growth/decline factor of the RM-rms relative to the RM-rms trend obtained for $1 \nG$ normalization. By predicting $\rm{RM}_{\rm{rms}}$ for various initial magnetic field strengths, we can then identify the largest field strength that is not excluded by observational data. This value is designated as an upper limit on the magnetic field strength.

The additional RM-rms trends using $0.1$ and $0.01 \nG$ normalization values for the km1 case are shown in the top panel of Figure~\ref{fig:RM-rmsEv-km1}. We also show $\rm{RM}_{\rm {rms}}$ for a range of $\rm{RM}_{\rm {growth}}$ factors in the same panel. As we see, the shape and amplitude of the estimated $\rm{RM}_{\rm {rms}}$ (Equation~(\ref{eq:RM-growth})) match well with the results from the $0.1$ and $0.01 \nG$ runs when $\rm{RM}_{\rm {growth}} = 0.1$ and  $\rm{RM}_{\rm {growth}} = 0.01$, respectively. Based on this analysis, for the km1 case, we obtain $0.75 \nG$ as the upper limit on the field strength. We expect that the PMF models with even larger coherence scales, along with the uniform model, exhibit the same behavior --- the RM-rms shape remaining unaffected by the lower ($<1 \nG$) normalization of the field. This would then lead to an upper limit of $0.15 \nG$ on the uniform field strength.

In the bottom panel of Figure~\ref{fig:RM-rmsEv-km1}, we show the $\mathrm{RM}_{\text{IGM}}$ trends for the whole LOS. In this case, we compute the RM maps without filtering out high-density regions.
As can be seen, RMs calculated for the entire LOS are higher than RMs extracted solely from the IGM. The difference between these two cases increases with redshift at lower redshifts ($z\lesssim 0.2$) and stays roughly the same ($\sim 1.5$) at higher redshifts. 
The same results are obtained for the rest of the PMF models.

\begin{figure}[htbp]
    \centering
    \includegraphics[width=7.0cm]{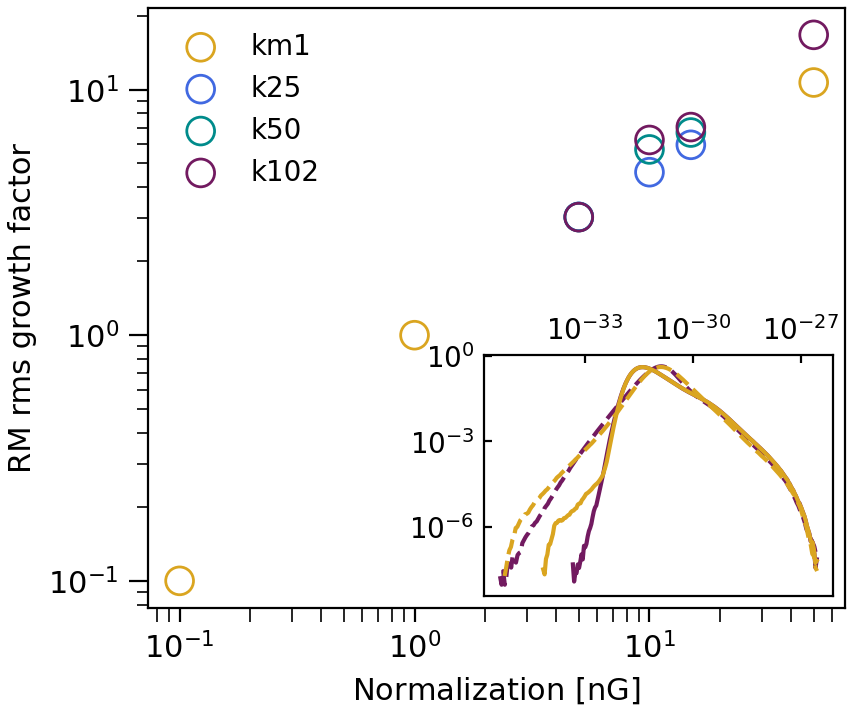}
    \caption{RM rms growth factors (relative to RM rms obtained for $1 \nG$ normalization) for the km1, k25, k50, and k102 cases. The inset panel shows the density PDFs with dashed lines PDFs from the $50 \nG$ normalizations (km1 and k102 cases). The units of the $x$-axis in the inset are in grams per cubic centimeter.
    } 
    \label{fig:RM_growth}
\end{figure}

In Figure~\ref{fig:RM_growth}, we provide the $\rm{RM}_{\rm {growth}}$ factors from the aforementioned additional runs. 
The growth factors for the km1, k25, and k102 cases are derived from our runs. For the k50 case, we interpolate the obtained growth factors from the k25 and k102 simulations. The shapes of RM-rms trends for the k25 and k102 cases remain mostly unaffected
by the initial magnetic field strength.\footnote{We emphasize that this is the
$50 \nG$ normalization where we see that the shape of the RM rms trend is
also affected (k102 model) and Equation~(\ref{eq:RM-growth}) is not valid anymore. However, this normalization is excluded by
observation data.}
The trends of rms growth factor are similar for k25, k50, and k102 models, although growth factors are larger for models characterized by smaller coherence scales. 
Smaller growth factors from larger ($> 1\nG$) field strengths possibly hint at a backreaction from such models on the structure formation processes, which may further influence the magnetic field growth and, consequently, RM evolution trends. In the inset panel of Figure~\ref{fig:RM_growth}, we show that the density PDF is affected by the initial normalization of the magnetic field. In particular, we see that in the k102 and km1 cases ($50\nG$ normalization), the shape of the density PDF changes at lower densities ($<10^{-30} \g \cm^{-3}$).
Finally, similarly to the case of the large-scale models, growth factors of smaller-scale fields are used to place upper limits on the strength of these PMFs. The obtained upper limits for the k25, k50, and k102 models are then $2.41, 7.85$, and $28.8 \nG$, respectively.

\subsection{Constraints on the magnetic field strength}
\label{subsec:constraints}

\begin{figure}[htbp]
    \centering
    \includegraphics[width=8.5cm]{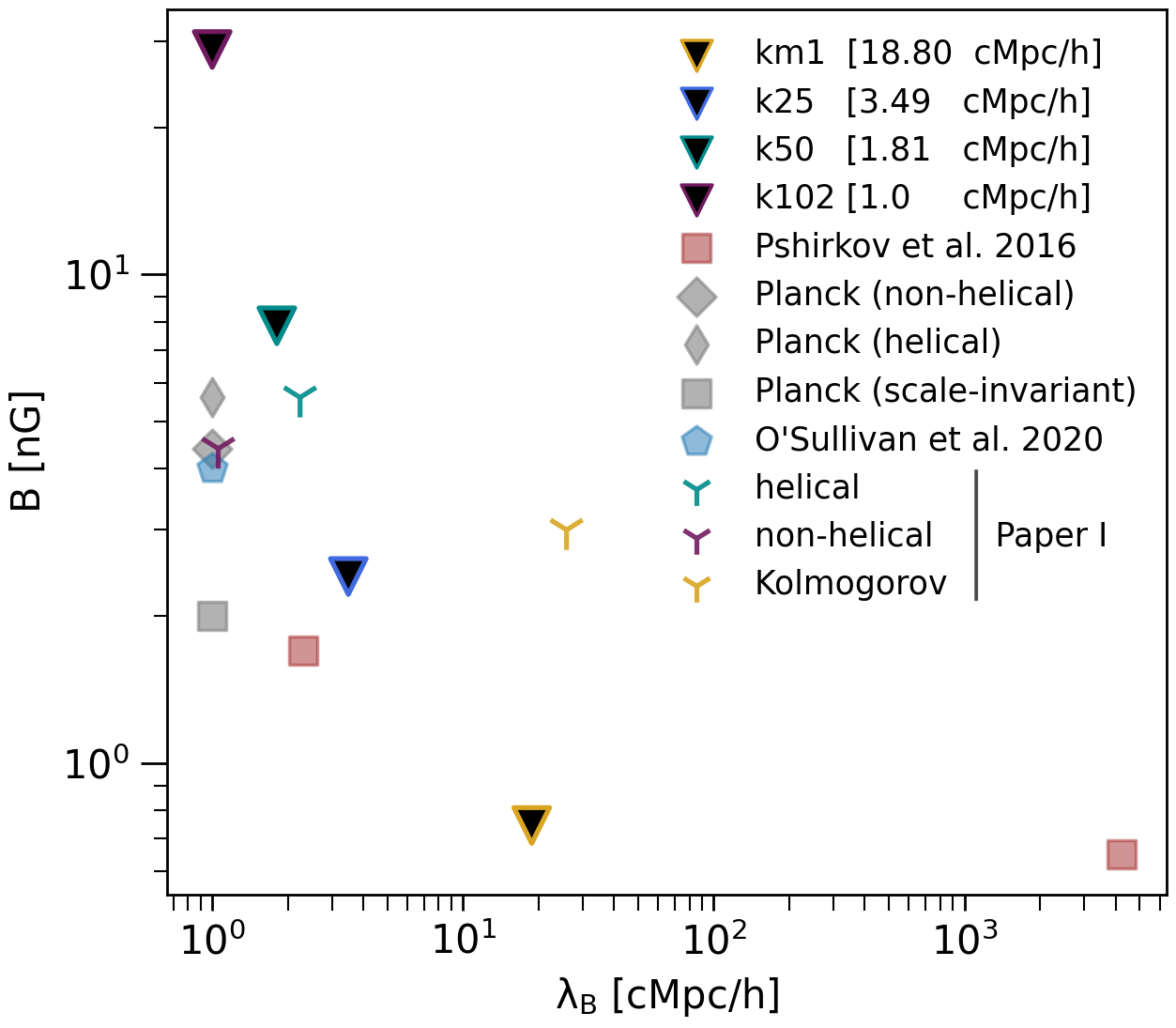}
    \caption{Constraints on the magnetic field strength when their different correlation lengths are considered. Upper limits derived in this work, and  
    in \cite{Plancketal2016} 
    are shown with filled black and gray markers, respectively.
We note that for \cite{Plancketal2016}, $\lambda_{\rm{B}}$ indicates the smoothing scale, and not the correlation length of the magnetic field.
    } 
    \label{fig:constraints}
\end{figure}

We summarize our analysis by presenting constraints on the PMF strengths in the $B - \lambda_{\text{B}}$ parameter space. Figure~\ref{fig:constraints} shows the upper limits on the strength of our PMF models along with the constraints on the magnetic field strength from other relevant studies \citep[see also Table~2 in][]{Amaraletal2021}. Even though it is common practice to combine and qualitatively visualize constraints from various work in the $B - \lambda_{\text{B}}$ parameter space \citep[see, e.g.,][for such examples]{NeronovSemikoz2009,DurrerNeronov2013,Brandenburgetal2017,Vachaspati2020,AlvesSav2021}, as emphasized by \cite{Vachaspati2020}, this plot should be treated with caution. The reason for this disclaimer mainly lies in the different definitions of the magnetic field strength and correlation length used in the literature. 
For instance, using various effects of PMFs (e.g., magnetically induced non-Gaussianities) on the CMB spectra, \citealt{Plancketal2016} puts constraints on the smoothed magnetic field strength; i.e., the magnetic field amplitude obtained by integrating the magnetic field power spectrum convolved with a Gaussian window function. In contrast, Faraday rotation measurements constrain the LOS (density)-averaged magnetic field (see \citealt{Vachaspati2020} for a summary of these definitions; see also \citealt{Ryuetal2008} who noticed differences between magnetic field amplitudes averaged using density or volume weights). Having this caveat in mind, 
in Figure~\ref{fig:constraints} we directly depict upper limits on the magnetic field strength from recent work. As we see from the figure, our constraints agree closely with those derived in other work. Similarly to the constraints derived from blazar spectra observations \citep{NeronovVovk2010}, our upper limits decrease with an increasing correlation length. It is interesting to notice that this trend is in good agreement
with the upper limit trends obtained in \cite{Pshirkovetal2016}; although, upper limit for the k102 model is higher than the upper limit derived in \cite{Plancketal2016}.
Constraints from the CMB analysis, combined with those derived in this work using the LoTSS survey, strongly disfavor fields with coherence scales $\lesssim 3.5 \cMpch$ for initial normalisations of $\nG$ (see also \cite{PaolettiFinelli2019} where $B_{1 \rm Mpc} \lesssim 0.04 \nG$ for stochastic small-scale models).
Finally, we see that the upper limits from this work are tighter
than the constraints obtained for our nonhelical and scale-invariant (Kolmogorov) fields in our previous work (Paper I); 
in Paper I, we used observation results from \cite{OSullivanetal2020} to place constraints on the strength of PMFs.
The upper limit of $0.15 \nG$ 
placed on the strength of the uniform model in this study is also tighter than our previous estimates, as well as can potentially be lower than what can be obtained from CMB analysis.

\section{Simulation and analysis uncertainty aspects}
\label{sec:NumAsp}
We generated a statistical sample of RM images by using different projection axes in Equation~(\ref{eq:RM}) for simulated data at different redshifts and by randomizing a center of the projected regions. 
This method does not ensure complete independence of the RM light-cone realizations;
i.e., the cosmic variance effects are not fully taken into account.
A future improvement in this direction is necessary to produce a larger (independent) statistical sample and account for (possibly) larger rms variations in the RM sky
and their dependence on the FOV.
We also note that, although our simulations seem to be converged, a future higher-resolution study might still be necessary to quantitatively assess the dependence of our results on resolution. In Figure~\ref{fig:RM-Ev-fittedLogx} of Appendix~\ref{app:RM-rms-fitting-resolution} we compare RM-rms trends obtained from our runs with trends from lower-resolution simulations. Based on this comparison, we do not expect the amplitude of RM-rms to be affected by more than a factor of $~1.5$ at higher redshifts in the higher-resolution simulation; see Figure~14 in Paper~I, where we showed that $\mathrm{RM}_{\text{IGM}}$ is converged at $z=0.02$. 
The dependence of amplification trends of stochastic small-scale models on the resolution has also been studied in Paper I; in Figure 13 of that paper, it is shown that up to $1.5$ times larger field strengths are expected in filaments from higher-resolution simulations.
Finally, we notice that both the filtering criteria for ``excluding-clusters'' regions and the methods used to subtract galaxy-cluster RMs from the total RM are expected to affect RM statistics \citep[][see also Figure 11 in Paper I]{AkahoriRyu2011}. \citealt{AkahoriRyu2011} showed that different cluster-RM subtraction methods can result in a factor of 1.5 difference too between RM-rms trends (see their Section 2.8 and Figure 5).

The RM-rms trends reported in \cite{Carrettietal2023} were derived from the RM catalog, which used Stokes $Q$ and $U$ data cubes from the LoTSS DR2 survey. The RMs were computed using the RM synthesis technique, enabling the translation of polarization measurements at different wavelengths into the Faraday dispersion spectrum (see also footnote on p. 2). A direct comparison with these observations is then only possible through modeling radio sources in simulations and obtaining RMs using a similar technique. While we caution the reader regarding this uncertainty --- specifically, the absence of radio-source modeling in our simulations and the RM synthesis technique --- we also emphasize that the light-cone technique employed in this study is suitable for tracking RMs induced solely by the foreground, magnetized IGM.

It should also be noted that RM observations with LOFAR are affected by depolarization \citep[see, e.g., Figure 17 in][]{OSullivanetal2023}, which leads to selection effects. The source sightlines detected in LOFAR are known to be biased against galaxy clusters because such an environment fully depolarizes background sources \citep{Carrettietal2022}. The source sightlines 
are only affected by
filaments and voids, which is a benefit for the detection of the $\mathrm{RM}_{\text{IGM}}$ term.\footnote{A forthcoming, more refined analysis (excluding sources with high  Galactic RM) of the LoTSS-survey by 
\citealt{Carrettietal2024}
has shown that the newly obtained RRM rms  exhibits a steeper evolution with redshift compared to the  trend reported by \citealt{Carrettietal2023}. However, the RRM values do not show significant changes at high redshifts ($z\gtrsim 1.25$); therefore, we expect that our upper limits on the PMF strength would only be mildly affected if we compared the simulation and observation analysis using this refined RRM(z) data.}

The $A_{rrm} (1+z)^{-2}$ term, which we included in our analysis, accounts for the astrophysical-origin RRM component, whose subtraction is needed to remain solely with the RM-IGM term and compare analysis with simulations of the cosmic-web magnetic field.  Uncertainty can arise 
from the redshift dependence of this term.
This is addressed in a forthcoming work \citep{Carrettietal2024}. For a discussion on  the uncertainty associated with the Galactic RM we refer the reader to \citet{Carrettietal2022} and \citet{Carrettietal2023}.

Various processes not considered in our simulations could influence the magnetization degree of the IGM and consequently, the PMF upper limits derived in this work. Examples of such processes include galactic winds and magnetized jets ejected from AGN. Although the efficiency of the magnetization of vast voids by these events is expected to be low \citep{Dolagetal2011,Becketal2013}, it remains unclear how far the Universe from the centers of galaxies can be magnetized by, e.g., powerful magnetized jets.
Recently, using National Radio Astronomy Observatory
Very Large Array Sky Survey data, \cite{Garciaetal2023} claimed that
RRM is sensitive to baryonic feedback, suggesting that magnetized bubbles (blown through the action of galactic winds and AGN) can solely account for the observed RRM trends. In \cite{BlunierNeronov2024}, on the other hand,  the LoTSS survey and analysis similar to the method employed by \citealt{Carrettietal2023} have been used to compare simulated and observed RRM evolution. Unlike the findings of \citealt{Garciaetal2023}, \cite{BlunierNeronov2024} concluded that magnetized bubbles overproduce RRM trends. Therefore, IGM magnetization resulting from such scenarios is not compatible with the magnetization of the Universe predicted by the RRM analysis from the  LoTSS survey. While as highlighted by \cite{Carrettietal2023} and \cite{OSullivanetal2023} (see their section 4.1), improvements can be made in RRM rms uncertainties with future surveys 
(SKA-LOW with an expected  higher source density, and LoTSS with larger coverage area and finer  resolution leading to a higher number of polarized sources)
and work, the aforementioned controversy can also hopefully be resolved through realistic modeling of AGN-driven ejecta (coupled with different PMF models) in our future simulations. 
Future surveys can also improve the estimate of the Galactic RM and improve the error associated with it.
 
%

\section{Conclusions and outlook}
\label{sec:Summ}

In this paper, we extended our previous work (Paper I) by (i) modeling PMFs in larger volumes to achieve FOVs relevant to LOFAR observations without the need for replicating simulated data, and (ii) employing a more realistic approach for comparing simulated RMs with the LoTSS data \citep{Carrettietal2023}. We used light cones \citep{yt-Turk2011} to generate a statistical sample of mock RM images at different redshift depths, enabling us to study the evolution of RM in the rarefied regions of the cosmic web. For the first time, we studied the dependence of $\mathrm{RM}_{\text{IGM}}$ trends on the initial strength and correlation length of PMFs while accounting for their cosmological MHD evolution during structure formation. Our analysis allows us to place constraints on PMF strengths, leading to the following conclusions:

\begin{itemize} 
\item[1.] The rms of $\mathrm{RM}_{\text{IGM}}$ exhibits redshift evolution for all PMF scenarios, and the trend for the  
large-scale-correlated PMF (with an initial scale-invariant spectrum)
is flattened at high redshifts ($z\gtrsim 1.5$);

\item[2.] A logarithmic fitting function provides a good fit for the simulated  $\mathrm{RM}_{\text{IGM}}$-rms trends. $\mathrm{RM}_{\text{IGM}}$-rms growth rates vary depending on the PMF model, with the fastest growth rates observed for our small-scale stochastic models (coherence scales: $3.5, 1.8$, and $1 \cMpch$);
rms slopes are similar for these models and show degeneracy
for the cases 
characterized by correlation lengths of $1.8$ and $1 \cMpch$.
In contrast, slopes of large- and small-scale PMF models are distinguishable;

\item[3.] 
The shape of the rms trends remains unaffected when derived for lower initial ($<1\nG$) normalizations of the field and for normalizations below or equal to $15 \nG$ in the 
stochastic model with the smallest coherence scale ($1 \cMpch$).
At higher redshifts, the shape of the RM-rms trend of the scale-invariant model better matches observations;

\item[4.] RMs are larger for PMFs with initial large-coherence scales. The average rms values throughout the $z=2-0$ redshift span are $6.3, 1.6, 0.5, 0.2 $, and $0.1~\text{rad}~ \text{m}^{-2}$, for our uniform, scale-invariant, and stochastic small-scale models with $3.5, 1.8$, and $1 \cMpch$ coherence scales, respectively; 

\item[5.] Our study,
using data from the recent analysis of the LoTSS survey \citep{Carrettietal2023},
places upper limits on the comoving magnetic field strength, with values 
$0.15, 0.75, 2.41, 7.85, 28.8 \nG$
for the uniform, scale-invariant, and stochastic models characterized by $3.5, 1.8$ and $1 \cMpch$ correlation lengths, respectively and thus, showing that upper limits are relaxed as the coherence scale of PMFs decreases. The obtained constraints are, qualitatively, in good agreement with the constraints from recent observations and from \cite{Pshirkovetal2016}, where upper limits on the magnetic field strength were derived using observation data along with semi-analytical estimates for RM. 
These constraints can be used in primordial magnetogenesis theories for predicting the strength and coherence scales of PMFs.
\end{itemize}

The constraints on the strength of PMFs obtained in this work can be used in primordial magnetogenesis theories for predicting the strength and coherence scales of PMFs, as well as in relic gravitational wave generation scenarios.
Our work went a step further in the creation of mock RM images. Hence, a joint effort of numerical modeling and observational advancements is underway in order to unveil the nature and origin of large-scale cosmic magnetism.
Our future study will complement the presented work by accounting for astrophysical magnetization, e.g., from powerful active galactic nuclei (AGN) jets to address controversy highlighted in recent literature \citep{Garciaetal2023,BlunierNeronov2024}. 
This future work will also enable us to investigate an excess RM contribution in the RM difference of physical and random pairs.
Furthermore, we will search for helical PMF imprints on the cross-correlation of RM and the degree of polarization of radio emission \citep{VolegovaStep2010,BranStep2014}.
Finally, we note that simulations presented in this work can also be used for other magnetic field-related problems in astrophysics and cosmology; examples include calculating axion-photon and graviton-photon conversion probabilities to ameliorate constraints on axion-like particle mass \citep[see, e.g.,][]{Matthewsetal2022} and the energy density of high-frequency gravitational waves \citep{Heetal2023}, respectively.

\section*{Acknowledgements}

S.M.\ is thankful to Ramkishor Sharma for his help in familiarizing us with the \textsc{Pencil Code}.
The authors appreciate useful discussions with and comments from Annalisa Bonafede, Sihao Cheng, Bryan Gaensler, Yutong He, Karsten Jedamzik, Michael Kachelriess, Chris Riseley, Tanmay Vachaspati,
and Jennifer West. 
We also acknowledge the scientific program ``Generation, evolution, and observations of cosmological magnetic fields'' held at the Bernoulli Center (2024) for hospitality and partial support.
S.M.\ and F.V.\ acknowledge financial support from 
Fondazione Cariplo and Fondazione CDP, through grant number Rif: 2022-2088 CUP J33C22004310003 for ``BREAKTHRU'' project.
S.M.\ was supported by the Shota Rustaveli National Science Foundation of Georgia, Georgian National Astrophysical Observatory, and Nordita. 
P.\ Dom\'inguez-Fern\'andez acknowledges the Future Faculty
Leaders Fellowship at the Center for Astrophysics, Harvard-Smithsonian. 
S.P.O acknowledges support from the Comunidad de Madrid Atracción de Talento program via grant 2022-T1/TIC-23797 and grant PID2023-146372OB-I00 funded by MICIU/AEI/10.13039/501100011033 and by ERDF, EU.
A.B.\ and T.K.\ acknowledge partial support from NASA Astrophysics Theory (ATP) 80NSSC22K0825 and NSF Astronomy and Astrophysics grant program (AAG) AST2307698, AST2408411 grants.
S.M. and F.V. acknowledge the Gauss Centre for Supercomputing e.V. (www.gauss-centre.eu) for supporting this project by providing computing time through the John von Neumann Institute for Computing (NIC) on the GCS Supercomputer JUWELS at J\"ulich Supercomputing Centre (JSC), under projects ``radgalicm2'', and the  the CINECA award  ``IsB28\_RADGALEO'' under the ISCRA initiative, for the availability of high-performance computing resources and support. Finally, we thank the anonymous reviewer for their helpful suggestions that improved the quality of this manuscript.

The computations described in this work were performed using the 
publicly available \texttt{Enzo} code (http://enzo-project.org), which is
the product of the collaborative efforts of many independent scientists from
numerous institutions around the world.  Their commitment to open science has helped make this work possible. 
The light-cone analysis was performed using the \texttt{yt\_astro\_analysis} extension \citep{Smithetal_2022} of the yt analysis toolkit \citep{yt-Turk2011}.
The simulations
presented in this work made use of computational resources on Norddeutscher Verbund f\"ur Hoch- und H\"ochstleistungsrechnen (HLRN; hhp00046) and JUWELS cluster at the Juelich Supercomputing Centre (JSC; under project ``Breakthru''); 
We also acknowledge the allocation of computing resources that was provided by the
Swedish National Allocations Committee at the Center for Parallel
Computers at the Royal Institute of Technology in Stockholm and
Link\"oping.

\section*{Data Availability}
The derived data supporting the findings of this study is freely available upon request. 
The simulation analysis data is available at \url{https://doi.org/10.5281/zenodo.14224378}.

{\large\em Software:} The source codes used for
the simulations of this study---\texttt{Enzo} \citep{2019JOSS....4.1636B} and
the {\sc Pencil Code} \citep{JOSS}---are freely available
online, at  \url{https://github.com/enzo-project/enzo-dev}
and \url{https://github.com/pencil-code/}. The yt analysis toolkit is also freely available \url{https://yt-project.org/}. \\

\typeout{} 
\bibliography{PMFsRM}{}
\bibliographystyle{aasjournal}

\appendix

\section{Magnetic field power spectrum}
\label{app:Bfield-PS}
\begin{figure}[htbp]
    \centering
    \includegraphics[width=7.3cm]{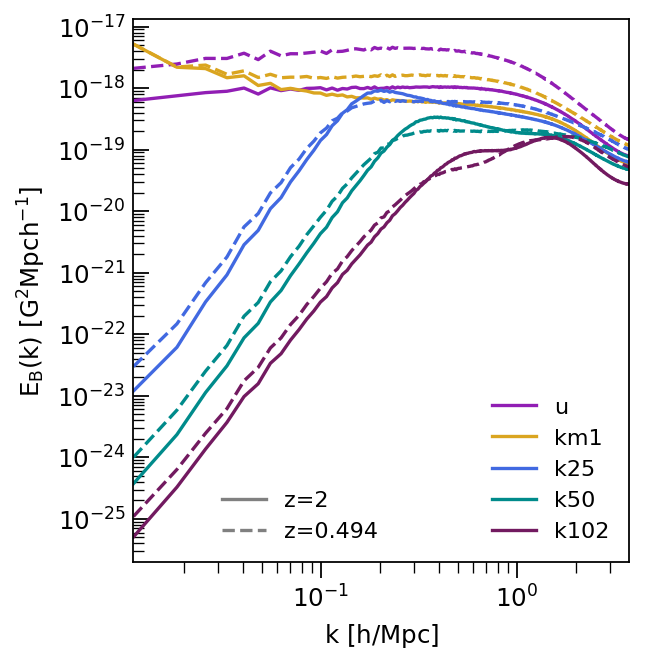}
   \caption{Magnetic field power spectrum at redshifts $z=2$ (solid lines) and $z=0.494$ (dashed lines) for different PMF models.}
    \label{fig:PS-redshift}
\end{figure}

In Figure~\ref{fig:PS-redshift}, we show magnetic field energy power spectra at $z=0.494$ and $z=2$ for all of our PMF models. In Figure~\ref{fig:projs}, we illustrated RM maps for the same redshift depths. From Figure~\ref{fig:PS-redshift} we observe that the power on the largest wavenumbers (our resolution limit) is nearly indistinguishable across different models, while on large scales, we see significant differences between the amplitudes of different PMFs. For the evolution of magnetic energy power spectra for models similar to those studied in this paper, we refer the reader to Section 4.3 of Paper I.
%

\section{Different light-cone realizations and the FOV dependence}
\label{app:PDFs-FOV-seeds}
\begin{figure}[htbp]
    \centering
    \includegraphics[width=6.7cm]{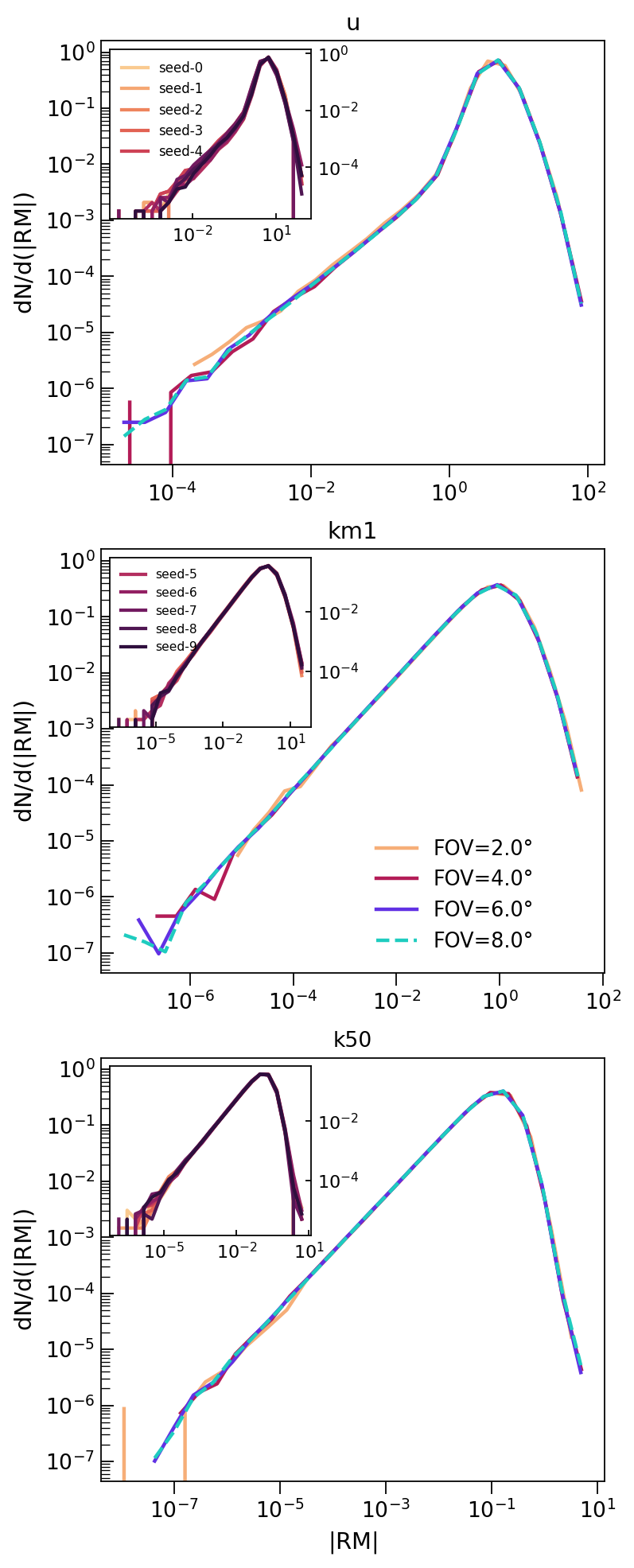}
   \caption{$|\textrm{RM}|$ PDF for different FOVs (solid and dashed lines) and different seedings (inset panels) for the uniform, stochastic scale-invariant, and stochastic k50 models. PDFs are obtained for $z=0.782$ depth. At this redshift, the proper width of our simulated volume is $112.73 \Mpc$; the given statistics for the \SI{6}{\degree}  and \SI{8}{\degree} FOVs correspond to $166.1 \Mpc$ and $221.5 \Mpc$, respectively. For the latter statistics, we replicated the simulation data.}
    \label{fig:RM-PDF-FOVdep}
\end{figure}

The dependence of the $|\textrm{RM}|$ PDF on the observer FOV at $z=0.782$ redshift depth is shown in Figure~\ref{fig:RM-PDF-FOVdep}. Statistics for each FOV are obtained by averaging $|\textrm{RM}|$ PDFs from different light-cone realizations. For each PMF model, we also show the PDFs for these different realizations
in the inset of the figure. As the figures show, the $|\textrm{RM}|$ PDF trends are the same from all realizations and for different FOVs.

\section{RM PDFs}
\label{app:PDFs-evolution}
%
\begin{figure*}[t]
    \centering
    \includegraphics[width=18.0cm]{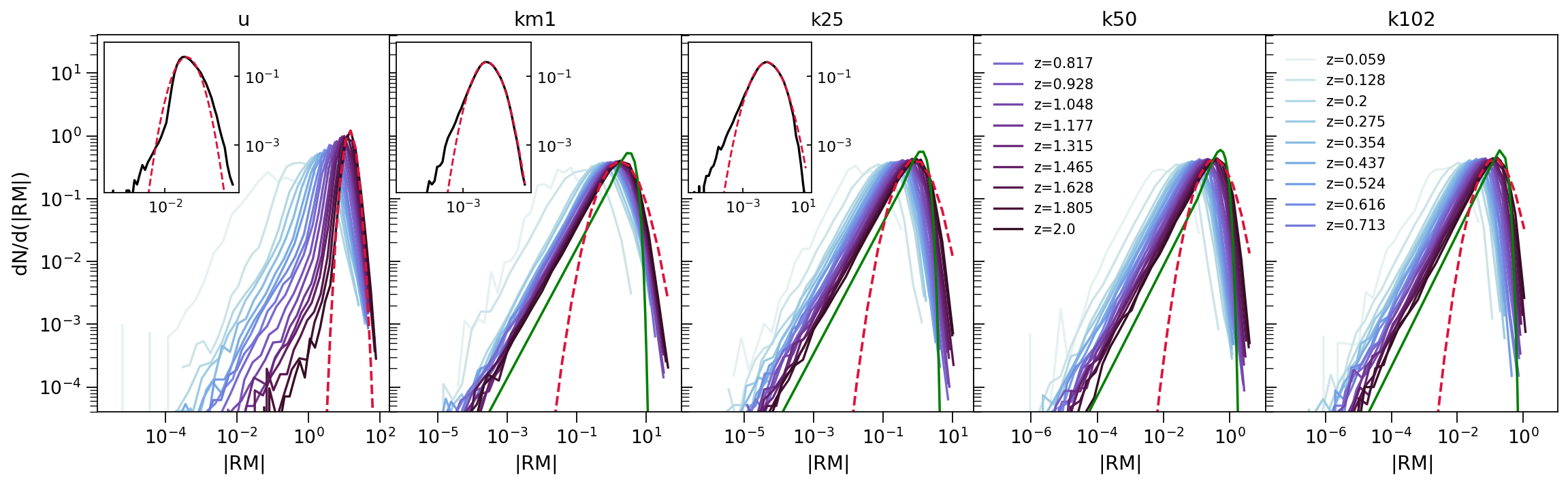}
   \caption{Redshift depth evolution of the $|\textrm{RM}|$ (excluding clusters) PDF for different PMF models with lognormal (red dashed) and Gaussian functions overplotted (green solid lines). The $|\textrm{RM}|$ PDF in the inset panel has been obtained for one ($z=2$) snapshot from our simulations.}
    \label{fig:RM-pdfEv}
\end{figure*}

The increasing trends of the RM rms evolution can also be understood in terms of a shift of the PDFs peak toward higher $|\textrm{RM}|$  values when considering longer LOSs. 
This is illustrated in Figure~\ref{fig:RM-pdfEv}. We see that, for all the models, more data points fall into the higher $|\textrm{RM}|$ bins as the distance between the source and the observer increases. 
We tried to fit the distribution using Gaussian and lognormal functions for $z=2$ redshift depth.  As we see, only the uniform model shows a rather good fit by the lognormal function, while the low-end tails of the PDFs in the stochastic cases follow a Gaussian distribution more closely. In the inset of the figure, we demonstrate that the $|\textrm{RM}|$  PDF for the simulation output at $z=2$ (i.e., only for one redshift snapshot) shows a better lognormal trend (high-end tails of the PDF) in the stochastic cases and has a power-law low-end tail. Finally, we also note that the PDF of the RM (not shown) is symmetric around zero in the stochastic cases and is skewed toward positive RM in the uniform model.

\section{RM-rms fitting and resolution test}
\label{app:RM-rms-fitting-resolution}
\begin{figure}[ht!]
    \centering
    \includegraphics[width=8.0cm]{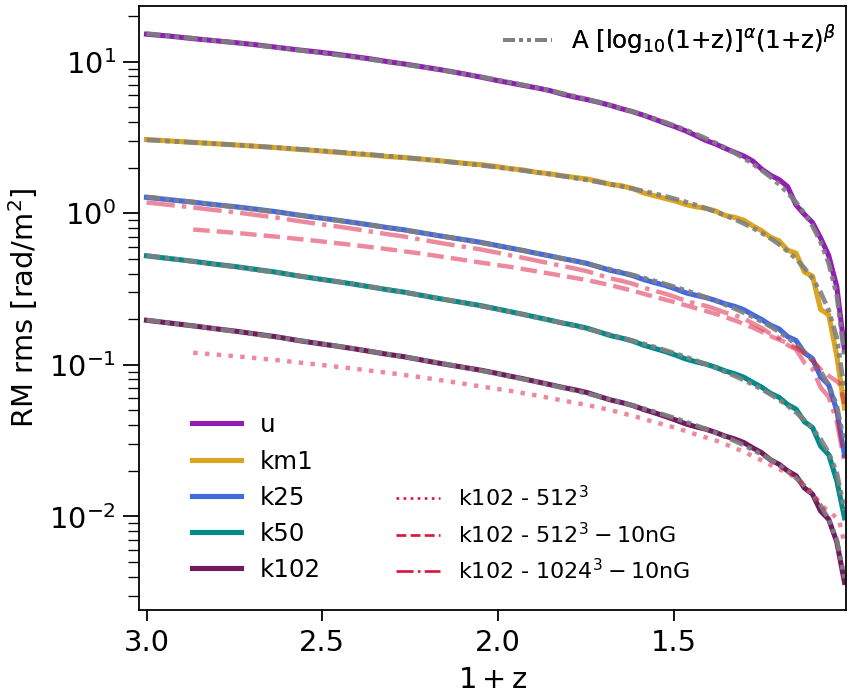}
   \caption{Evolution of $\mathrm{RM}_{\text{IGM}}$ for all PMF models and corresponding logarithmic fits (gray dashed-dotted lines). Red dotted, dashed, and dash-dotted lines show the RM-rms trends of the k102 model from $512^3$,  $512^3$-$10\nG$ normalization runs and the $1024^3$-$10\nG$ simulation, respectively.}
    \label{fig:RM-Ev-fittedLogx}
\end{figure}

In Figure~\ref{fig:RM-Ev-fittedLogx}, we illustrate the fitting of RM-rms trends using Equation~(\ref{eq:fittingFun}). As we see, this function fits RM-rms trends of different PMF models well at all redshifts and can thus be used in future observation surveys to mimic trends from our models. In the same figure, we also show that RM trends are mostly converged at low redshifts for both $1 \nG$ and $10\nG$ normalizations of the k102 model. 

\listofchanges
\end{document}